\documentclass[aps,prx,superscriptaddress,amsmath,amssymb,twocolumn,showpacs,floatfix,reprint]{revtex4-2}
\usepackage[T1]{fontenc}
\usepackage{amsmath}
\usepackage{amssymb}
\usepackage{algorithm}
\usepackage{algpseudocode}
\usepackage[english]{babel}
\usepackage{booktabs}
\usepackage{graphicx}
\usepackage[colorlinks=true,citecolor=blue,linkcolor=blue]{hyperref}
\usepackage{multirow}
\usepackage{newtxmath}
\usepackage{newtxtext}

\newcommand{\s}{\mathbf{s}}

\definecolor{C0}{HTML}{1f77b4}
\definecolor{C1}{HTML}{ff7f0e}
\definecolor{C2}{HTML}{2ca02c}
\definecolor{C3}{HTML}{d62728}
\definecolor{C4}{HTML}{9467bd}
\definecolor{C5}{HTML}{8c564b}

\begin{document}
% \title{Enhancing PEPS Sampling with Single-layer Autoregressive Row Update}
% \title{Efficient Sampling of Projected Entangled-Pair States in Rugged Landscapes via Single-layer Autoregressive Row Update}
% \title{Efficient Sampling of Projected Entangled-Pair States in Rugged Landscapes using Single-layer Autoregressive Row Update}
\title{Variational Monte Carlo (VMC) with row-update Projected Entangled-Pair States (PEPS) and its applications in quantum spin glasses}

\author{Tao Chen}%
\affiliation{
Hefei National Laboratory for Physical Sciences at the Microscale and Department of Modern Physics, University of Science and Technology of China, Hefei 230026, China}
\affiliation{Hefei National Laboratory, University of Science and Technology of China, Hefei 230088, China}

\author{Jing Liu}
\affiliation{School of Physical Science and Technology, Beijing University of Posts and Telecommunications, Beijing 100876, China}
\affiliation{
CAS Key Laboratory for Theoretical Physics, Institute of Theoretical Physics,
 Chinese Academy of Sciences, Beijing 100190, China}

\author{Yantao Wu}
\thanks{yantaow@iphy.ac.cn}
\affiliation{Institute of Physics, Chinese Academy of Sciences, Beijing 100190, China}

\author{Pan Zhang}%
\thanks{panzhang@itp.ac.cn}
\affiliation{Hefei National Laboratory, University of Science and Technology of China, Hefei 230088, China}
\affiliation{
CAS Key Laboratory for Theoretical Physics, Institute of Theoretical Physics,
 Chinese Academy of Sciences, Beijing 100190, China}
\affiliation{
School of Fundamental Physics and Mathematical Sciences, Hangzhou Institute for Advanced Study, UCAS, Hangzhou 310024, China
}
\affiliation{School of Physical Sciences, University of Chinese Academy of Sciences, Beijing 100049, China}

\author{Youjin Deng}%
\thanks{yjdeng@ustc.edu.cn}
\affiliation{
Hefei National Laboratory for Physical Sciences at the Microscale and Department of Modern Physics, University of Science and Technology of China, Hefei 230026, China}
\affiliation{Hefei National Laboratory, University of Science and Technology of China, Hefei 230088, China}

\date{\today}

\begin{abstract}
Solving the quantum many-body ground state problem remains a central challenge in computational physics. In this context, the Variational Monte Carlo (VMC) framework based on Projected Entangled Pair States (PEPS) has witnessed rapid development, establishing itself as a vital approach for investigating strongly correlated two-dimensional systems. However, standard PEPS-VMC algorithms predominantly rely on sequential local updates. This conventional approach often suffers from slow convergence and critical slowing down, particularly in the vicinity of phase transitions or within frustrated landscapes.
To address these limitations, we propose an efficient autoregressive row-wise sampling algorithm for PEPS that enables direct, rejection-free sampling via single-layer contractions. By utilizing autoregressive single-layer row updates to generate collective, non-local configuration proposals, our method significantly reduces temporal correlations compared to local Metropolis moves. We benchmark the algorithm on the two-dimensional transverse-field Ising model and the quantum spin glass. Our results demonstrate that the row-wise scheme effectively mitigates critical slowing down near the Ising critical point. Furthermore, in the rugged landscape of the quantum spin glass, it yields improved optimization stability and lower ground-state energies. These findings indicate that single-layer autoregressive row updates provide a flexible and robust improvement to local PEPS-VMC sampling and may serve as a basis for more advanced sampling schemes.
\end{abstract}

\maketitle

\emph{Introduction.}---The theoretical treatment of quantum many-body systems remains a central challenge in condensed matter physics due to the exponential scaling of the Hilbert space dimension. This difficulty has driven the development of tensor network methods, which efficiently encode the low-energy physics of complex quantum systems by exploiting their local entanglement structure~\cite{PhysRevLett.69.2863,PhysRevLett.93.040502,PhysRevLett.93.227205,PhysRevLett.96.220601,PhysRevLett.99.220405}. 
For two-dimensional (2D) models, projected entangled pair states (PEPS)~\cite{PhysRevLett.101.250602,PhysRevLett.101.090603,PhysRevB.90.064425,PhysRevB.95.195154,PhysRevB.99.195153,PhysRevLett.124.037201,PhysRevB.103.235155} have emerged as the natural generalization of one-dimensional matrix product states (MPS)~\cite{PhysRevLett.69.2863,SCHOLLWOCK201196,PhysRevB.48.10345}, explicitly designed to capture the area-law scaling of entanglement entropy characteristic of local Hamiltonian ground states.
Despite their theoretical appeal, the practical power of PEPS is constrained by a severe computational bottleneck, in stark contrast to the success of MPS in one dimension. Accurate ground-state simulations using conventional double-layer contraction schemes entail a computational cost scaling as $O(D^{10})-O(D^{13})$ on square lattices, with memory cost scaling as $O(D^{8})$~\cite{PhysRevLett.99.220405,PhysRevLett.101.250602,PhysRevB.80.094403,PhysRevLett.103.160601,PhysRevB.86.045139,PhysRevB.90.064425,PhysRevB.92.201111}, where $D$ is the bond dimension determining the representation ability of the PEPS. This severely limits the practical application of PEPS for 2D systems. 
To circumvent these prohibitive costs, the combination of PEPS with Variational Monte Carlo (VMC) offers a rigorous and efficient alternative~\cite{PhysRevB.95.195154,PhysRevB.103.235155,PhysRevLett.134.256502}. In the PEPS-VMC framework, contractions are reduced to a single layer, lowering the complexity to $O(D^6)-O(D^8)$ in time and $O(D^4)$ in memory. This reduction enables gradient-based optimization and the evaluation of observables via massively parallelized sampling~\cite{PhysRevLett.99.220602,PhysRevLett.100.040501}.

% ---
% However, the power of 2D TNSs, specifically PEPS, is seriously hindered by their complexity, in stark contrast to the huge success of 1D MPS. The primary challenge is the extremely expensive computational cost for accurate simulations. In conventional double-layer contraction schemes, this cost scales at least as $O(D^{10})$ on square lattices, with memory cost scaling as $O(D^{8})$~\cite{PhysRevLett.99.220405,PhysRevLett.101.250602,PhysRevB.80.094403,PhysRevLett.103.160601,PhysRevB.86.045139,PhysRevB.90.064425,PhysRevB.92.201111}, where D is the bond dimension determining the representation ability of the PEPS. This severely limits the practical application of PEPS for 2D systems. Among various optimization algorithms, the combination of PEPS with Variational Monte Carlo (VMC) sampling provides an elegant framework to overcome these difficulties~\cite{PhysRevB.95.195154,PhysRevB.103.235155,PhysRevLett.134.256502}. In the PEPS-VMC scheme, one deals only with single-layer tensor networks, reducing the computational cost to $O(D^6)$ and memory cost to $O(D^4)$. This approach allows for gradient-based optimization, and physical observables can be evaluated efficiently via MC sampling with massive parallelization~\cite{PhysRevLett.99.220602,PhysRevLett.100.040501}. 

However, the efficacy of the PEPS-VMC framework relies critically on the efficiency of the sampling algorithm used to generate configurations $\s$ from the probability distribution $P(\s) \propto |\Psi(\s)|^2$. Standard approaches utilize Markov Chain Monte Carlo (MCMC) methods, such as the Metropolis-Hastings algorithm~\cite{Metropolis1953EquationOS}, which propose local, single-spin updates.
While conceptually simple, this approach faced a severe bottleneck: calculating the acceptance probability for a random site required a computationally expensive contraction of the full tensor environment. An important advance by Liu \emph{et al.}~\cite{PhysRevB.95.195154,PhysRevB.103.235155} introduced a deterministic sequential update scheme (e.g., following a ``zigzag'' path) that allows for extensive reuse of environment tensors. 
The detailed balance of the sequential update scheme was later proved in Ref. \cite{wu2025algorithms}.
This innovation substantially reduced the per-sweep complexity, making high-bond-dimension PEPS simulations feasible. Nevertheless, sampling dynamics driven by local or sequential updates face intrinsic limitations in challenging physical regimes. Near continuous phase transitions, local updates suffer from critical slowing down~\cite{PhysRev.145.224}, resulting in diverging autocorrelation times. Furthermore, in systems with frustrated interactions, such as quantum spin glasses~\cite{Edwards1975,binder1986}, the rugged energy landscape impedes ergodicity; local updates frequently become trapped in local minima, exploring restricted regions of the phase space. These issues compromise the stability of the variational optimization and amplify statistical noise in observables. An alternative is the direct sampling approach~\cite{SR_3}, which generates whole configurations from conditional probabilities as MCMC proposals. However, this typically requires a double-layer contraction of the PEPS, incurring a significantly higher computational overhead that limits its scalability.

% Nevertheless, the sampling dynamics generated by local or sequential updates can remain inefficient in challenging physical regimes. Near criticality, local updates are affected by critical slowing down~\cite{PhysRev.145.224}, leading to increased autocorrelations between samples. In systems with frustrated interactions, such as quantum spin glasses (QSG)~\cite{Edwards1975,binder1986}, the rugged structure of the configuration space further impedes sampling efficiency, as local updates may spend long times exploring restricted regions of phase space. While these issues do not necessarily invalidate the sampling procedure, they can reduce the stability of variational optimization and increase statistical fluctuations in measured observables.

In this work, we explore a non-local sampling strategy for PEPS-VMC based on autoregressive row-wise updates. While the update remains restricted to a single row at a time rather than the full lattice, it goes beyond strictly local spin flips by enabling collective, row-wise configuration changes. 
Leveraging the efficient environment construction developed for sequential updates, we compute exact conditional probabilities for entire rows of spins, enabling collective Monte Carlo moves that extend beyond strictly local updates. Crucially, this approach maintains the same computational complexity as standard local sampling strategies. We benchmark this technique on the transverse-field Ising model and a quantum spin glass in two dimensions (2D). Our results demonstrate that this proposed sampling scheme significantly suppresses temporal correlations and enhances the stability of variational optimization, particularly in systems with complex energy landscapes. These findings demonstrate autoregressive row-wise sampling as a useful algorithmic enhancement for existing PEPS-VMC methodologies.

\begin{figure}
    \centering
    \includegraphics[width=1.0\linewidth]{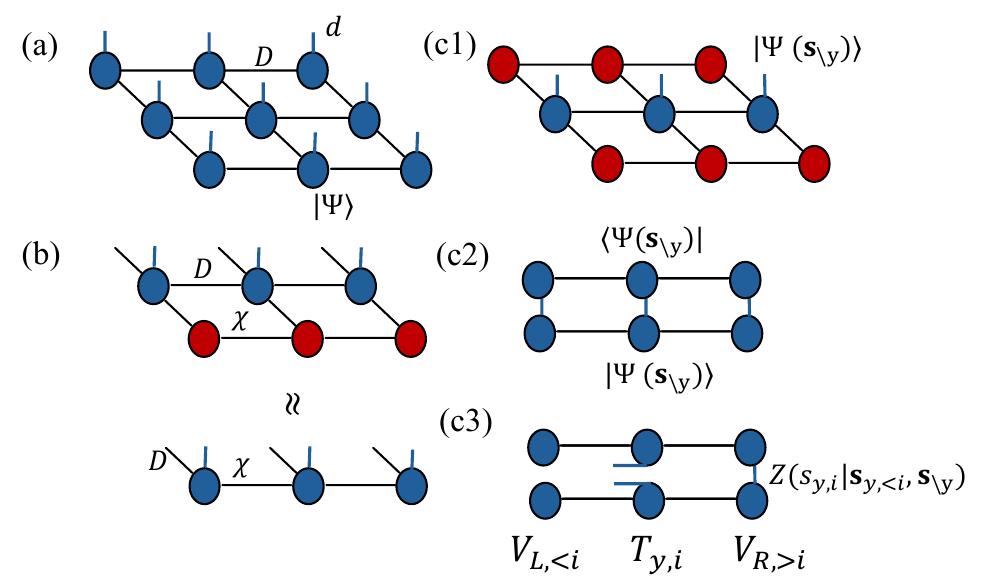}
    \caption{Schematic illustration of the single-layer autoregressive row-update algorithm. (a) The projected entangled pair state (PEPS) ansatz representing the quantum state $|\Psi\rangle$. (b) Construction of the effective environment for the target row $y$. The tensor network is partitioned, and the sub-networks above and below row $y$ are contracted into boundary matrix product states (MPS) with bond dimension $\chi$. (c1)-(c3) Sequential autoregressive sampling along row $y$. The conditional probability for spin $s_{y,i}$ is derived by contracting the local tensor with the environment comprising the boundary MPS and the previously sampled spins $\s_{y,<i}$. This procedure yields the marginal partition function (rightmost panel), enabling direct sampling without Metropolis rejection.}
    \label{fig:PEPS}
\end{figure}

\emph{Method.}---We consider a PEPS ansatz on a 2D lattice, and wave function coefficient $\Psi(\s)$ corresponding to a quantum state $|\Psi\rangle = \sum_{\s} \Psi(\boldsymbol{s})|\s\rangle$ is obtained via the full contraction of the sample-sliced single layer tensor network: $\Psi(\s) = \text{Tr}(\prod_i T[i]^{p=\s_i})$, as depicted in Fig.~\ref{fig:PEPS}(a).
Here, $\s = \{s_1, s_2, \dots, s_N\}$ denotes the global spin configuration. 
Each site $i$ is assigned a rank-5 tensor $T[i]^p_{lrud}$, where the physical index $p$ represents the local degree of freedom, and the four virtual indices $(l, r, u, d)$ connect to neighboring tensors.
The dimension of these virtual indices, denoted as the bond dimension $D$, controls the expressive power of the PEPS ansatz.
Within the VMC framework~\cite{Becca_Sorella_2017}, we sample configurations from the probability distribution $P(\s) \propto |\Psi(\s)|^2$ to estimate the energy expectation value 
\begin{equation}
    E_{\text{tot}} = \langle \Psi | H | \Psi \rangle / \langle \Psi | \Psi \rangle = \frac{1}{Z}\sum_{\s} |\Psi(\s)|^2E_{\text{loc}}(\s),
\end{equation}
where the local energy term $E_{\text{loc}}(\s)$ is defined as
\begin{equation}
    E_{\text{loc}}(\s) = \sum_{\s'}\frac{\Psi(\s')}{\Psi(\s)}\langle \s'|H|\s \rangle.
\end{equation}
The tensors $T[i]$ are optimized by minimizing $E_{\text{tot}}$ via stochastic gradient descent. The gradient of the energy with respect to the tensor elements is given by, assuming real parameters,
\begin{equation}
    \frac{\partial E_{\text{tot}}}{\partial T[i]^p_{lrud}} = \langle O[i]^p_{lrud}(\s)E_{\text{loc}}(\s)\rangle -\langle O[i]_{lrud}^{p}(\s)\rangle\langle E_{\text{loc}}(\s)\rangle,
\end{equation}
where $\langle \cdots \rangle$ denotes the statistical average over the Monte Carlo samples. 
The term $O[i]^p_{lrud}(\s)$ is defined as the logarithmic derivative of the wave function:
\begin{equation}
    O[i]^p_{lrud}(\s) = \frac{1}{\Psi(\s)}\frac{\partial \Psi(\s)}{\partial T[i]^p_{lrud}} = 
    \begin{cases} o[i]_{lrud}(\s) & \text{if } p = \s_i\\
    0 & \text{otherwise}
    \end{cases}
\end{equation}
Here, $o[i]_{lrud}(\s)$ is formed via contraction of whole single-layer network excluding the tensor at site $i$ for a fixed configuration $\s$:
\begin{equation}
    o[i](\s) = \frac{1}{\Psi(\s)}\prod_{j\not= i} T[j]^{p=\s_j}.
\end{equation}
Consequently, both the energy and its gradients can be efficiently evaluated via Monte Carlo sampling to optimize the PEPS ansatz.
Standard PEPS-VMC employs the Metropolis-Hastings algorithm: a site $k$ is selected, and a spin flip $s_k \to s'_k$ is proposed, which is accepted with probability $P(\s \to \s') = \min(1, |\Psi(\s')|^2/|\Psi(\s)|^2)$. While random site selection is common, a deterministic sequential sweep was introduced in Ref.~\cite{PhysRevB.95.195154,PhysRevB.103.235155,wu2025algorithms}, yielding significant improvements in computational efficiency and thermalization. Nevertheless, the sampling dynamics is still based on local single-spin updates, which can result in long autocorrelation times when collective degrees of freedom dominate or when the configuration space exhibits a complex energy landscape. This observation motivates the exploration of non-local update strategies that extend beyond strictly local Metropolis moves.
% and a new configuration $\s'$ is proposed by flipping its spin ($s_k \to s'_k$). The proposal is accepted with probability:$P(\s \to \s') = \min(1, |\Psi(\s')|^2/|\Psi(\s)|^2)$. A significant improvement to this process, pioneered by Liu \emph{et al.}~\cite{PhysRevB.95.195154,PhysRevB.103.235155}, was the move from random site selection to a systematic, sequential scan (i.e., updating spins one by one from beginning to end). This deterministic sweep strategy was shown to dramatically improve computational efficiency and the decorrelation of samples. 

We introduce a non-local, collective update scheme that targets an entire row of spins.
% \todo{the notation is not optimal, because we use $s_1, \cdots, s_N$ above, any better notation?}
While sampling the spins of the $y$-th row, $\s_y =\{s_{y,1}, s_{y,2}, \dots, s_{y,L}\}$, directly from the joint conditional distribution $P(\s_y | \s_{\setminus y})$ is generally intractable (where $\s_{\setminus y}$ denotes all spins excluding the $y$-th row), the chain rule of probability allows for an exact decomposition into tractable conditional probabilities:
\begin{equation}
    P(\s_y | \s_{\setminus y}) = \prod_{i=1}^L P(s_{y,i} | s_{y,1}, \dots, s_{y,i-1}, \s_{\setminus y}).
\end{equation}
This decomposition enables the direct, exact sampling of the row sequence without the need for Metropolis rejection steps.
% \todo{should reorganize the next sentence...}The computation of these conditional probabilities is enabled by the tensor network structure. 
As illustrated in Fig.~\ref{fig:PEPS}(b), the PEPS tensors above and below the target row $y$ are contracted into two boundary MPS with a truncation bond dimension $\chi$.
% for a chosen row $y$, the PEPS tensors in the regions above and below are approximately contracted into two boundary MPS.  This contraction is controlled by a truncation bond dimension $\chi$, which limits the size of these environment MPS. 
These MPS form the environment for $y$-row, leading to the tensor network structure shown in Fig.~\ref{fig:PEPS}(c1). 
The sampling then proceeds autoregressively from $i=1$ to $L$, and the conditional probability for site $i$ is given by 
\begin{equation}
    P(s_{y,i} | \s_{y,<i}, \s_{\setminus y}) = Z(s_{y,i}|\s_{y,<i}, \s_{\setminus y})/ \sum_{s_{y,i}} Z(s_{y,i}|\s_{y,<i}, \s_{\setminus y}),
\end{equation}
where $Z(s_{y,i}|\s_{y,<i}, \s_{\setminus y})$ is the marginal partition function computed by contracting the network with fixed spins $\s_{\setminus y}$ and $\s_{y,\le i}$, while tracing over the unsampled spins $\s_{y,>i}$. This autoregressive approach shares conceptual similarities with recent advances in Tensor Network Monte Carlo for classical systems~\cite{SciPostPhys.14.5.123,PhysRevB.111.094201,chen2025tensornetworkmarkovchain,chen2025batchtnmc}.
A detailed description of the row-sampling is given in Supplemental Material.

The computational cost of a row update is comparable to that of a full sequential local-update sweep, as both rely on similar single-layer PEPS contractions and environment reuse.
Despite this comparable cost, the row update substantially reduces temporal correlations among Monte Carlo samples and promotes more global exploration of the configuration space.
However, for systems with rugged energy landscapes, such as quantum spin glasses, a pure row-update scheme may still become trapped in specific configurations. 
To address this, we employ a hybrid sampling strategy. In each Monte Carlo step, we first perform an autoregressive row update, where configurations for all rows are updated sequentially from one boundary to the other. This is immediately followed by a full Metropolis sweep over all individual sites of the lattice.
This hybrid approach combines the complementary strengths of the two methods: the row update enables large-scale exploration between distinct energy basins, while the subsequent Metropolis sweep facilitates efficient local relaxation within a basin.

\emph{Results.}---To quantify the efficacy of the proposed autoregressive sampling strategy in both MC equilibration and variational optimization, we benchmark three sampling schemes—conventional Metropolis updates, autoregressive row updates, and the hybrid strategy—on two representative two-dimensional quantum models: the transverse-field Ising model (TFIM)~\cite{TFIM_deng} and the quantum spin glass (QSG)~\cite{Bernaschi2024}.
% We consider the unified Hamiltonian
The physics of these models is captured by the unified Hamiltonian:
\begin{equation}
H = - \sum_{\langle i,j \rangle} J_{ij} \sigma_i^z \sigma_j^z - \Gamma \sum_i \sigma_i^x ,
\label{eq:hamiltonian}
\end{equation}
where $\sigma_i^\alpha$ are Pauli matrices. For the TFIM, the ferromagnetic coupling is uniform, $J_{ij}=J>0$. For the QSG, the couplings $J_{ij}$ are drawn independently from a bimodal distribution $J_{ij}=\pm J$ with equal probability.

\begin{figure}
    \centering
    \includegraphics[width=1.0\linewidth]{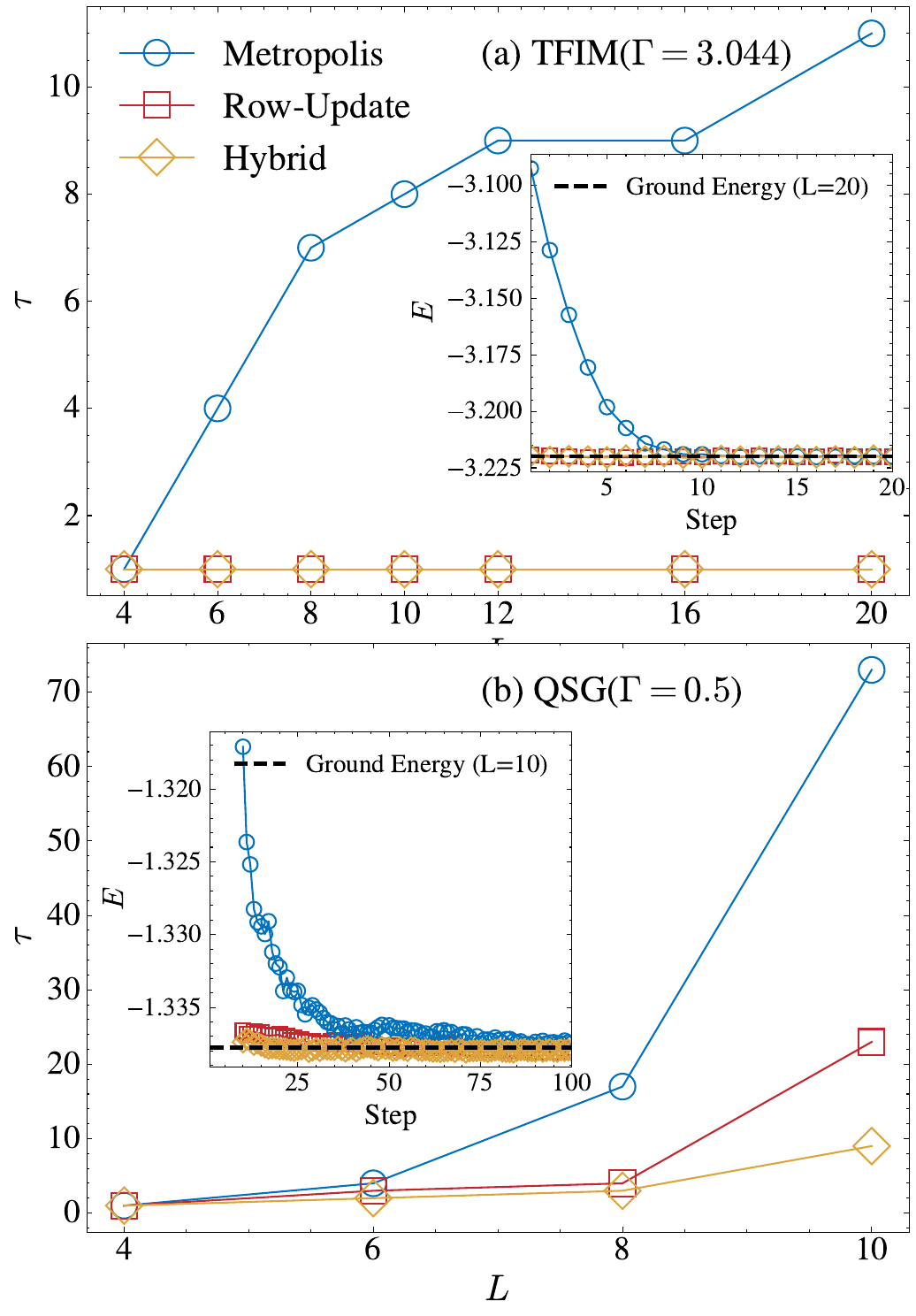}
    \caption{Markov chain equilibration dynamics initialized from a random high-energy state. The panels display the relaxation of the variational energy toward the pre-optimized ground state for (a) the 2D TFIM at the critical point $\Gamma_c \approx 3.044J$ and (b) the 2D QSG at $\Gamma=1.0J$. Main panels: Equilibration step $\tau$ as a function of system size $L$. Insets: Energy trajectories for the largest system sizes ($L=20$ for TFIM, $L=10$ for QSG). }
    \label{fig:equilibration}
\end{figure}

We first examine the Monte Carlo equilibration toward a fixed target distribution.
For this purpose, we pre-optimize a PEPS with bond dimension $D=3$ for each model, which defines a reference probability landscape $P(\s)=|\Psi_g(\s)|^2$.
Fixing the PEPS network parameters, we initialize 1000 independent Markov chains from random high-energy configurations and monitor the relaxation of the average energy.
The equilibration time $\tau$ is then defined as the number of Monte Carlo iterations required for the energy to converge within $10^{-3}$ of the reference PEPS expectation value.
All environment contractions are performed using a uniform MPS bond dimension $\chi=9$.
Figure~\ref{fig:equilibration} illustrates the sacling of the equilibration step $\tau$ with system size. For the TFIM at criticality ($\Gamma_c \approx 3.044J$) [Fig.~\ref{fig:equilibration}(a)], the Metropolis sampler exhibits characteristic critical slowing down, with the equilibration step following a power law.
In stark contrast, both the autoregressive and the hybrid strategies converge in a single iteration regardless of the system size, indicating that the row-update mechanism effectively generates highly decorrelated samples in the critical regime. It should be noted, however, that while this qualitative observation highlights the improved decorrelation efficiency of row-wise updates, the practical impact during an actual variational optimization for the TFIM is less pronounced. In the standard TFIM VMC optimization, the energy landscape is relatively smooth, and all sampling schemes—including conventional Metropolis updates—converge to the ground state with comparable speed and accuracy.
% for all tested system sizes, indicating near-perfect decorrelation and robust ergodicity.
The QSG at $\Gamma=0.5J$ presents a more challenging test [Fig.~\ref{fig:equilibration}(b)],
% The situation is more challenging for the QSG at $\Gamma=0.5J$ [Fig.~\ref{fig:equilibration}(b)], 
where the rugged free-energy landscape leads to exponential barriers. While all methods eventually exhibit an exponential scaling of thermalization, the non-local strategies suppress the prefactor of this growth. The hybrid strategy yields the fastest relaxation [Fig.~\ref{fig:equilibration} insets], demonstrating that combining global autoregressive moves with local refinement significantly enhances ergodicity in glassy landscapes.
% growth of equilibration steps for all methods. 
% Nevertheless, the non-local sampling strategies dramatically reduce the prefactor of this growth. The row-update method substantially outperforms Metropolis sampling, while the hybrid strategy further improves equilibration, achieving the smallest exponential growth rate. The insets of Fig.~\ref{fig:equilibration} show representative energy trajectories for the largest system sizes, illustrating that the hybrid method consistently relaxes faster and more reliably than the other approaches. These results demonstrate that autoregressive row sampling effectively mitigates critical slowing down in near-critical systems and significantly enhances exploration efficiency in glassy energy landscapes.

\begin{figure}
    \centering
    \includegraphics[width=1.0\linewidth]{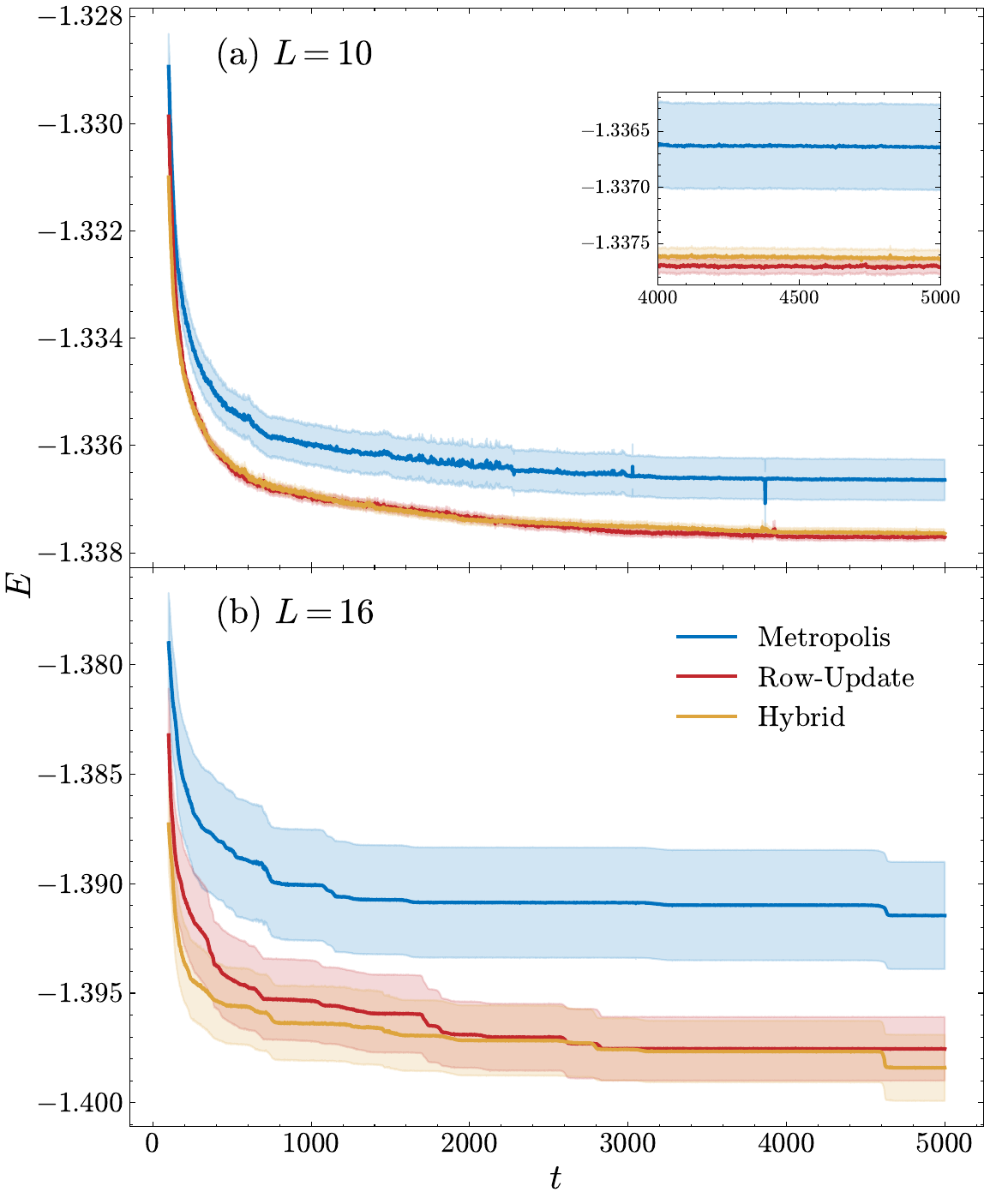}
    \caption{Convergence analysis of variational optimization across independent realizations. The evolution of the mean variational energy and its standard deviation (shaded regions) is shown for (a) $L=10$ (100 independent seeds) and (b) $L=16$ (20 independent seeds) QSG lattices. Both the pure Autoregressive Row-Update and the Hybrid methods demonstrate superior convergence properties, achieving significantly lower asymptotic energies with minimal variance compared to the Metropolis benchmark, which exhibits strong initialization dependence.}
    % Statistical performance across multiple seeds.
    % Performance statistics of the optimization algorithms over multiple independent runs with different random seeds. The plots show the mean energy and standard deviation (error bars) at each optimization step. (a) $L=10$ lattice with 100 independent seeds. (b) $L=16$ lattice with 20 independent seeds. In both cases, the Autoregressive Row-Update and Hybrid methods achieve significantly lower asymptotic energies and smaller variances compared to the Metropolis benchmark, indicating superior robustness against initialization bias.}
    \label{fig:spin_glass_energy}
\end{figure}

% We now turn to the central question of variational robustness in the presence of a highly frustrated energy landscape. In this setting, sampling inefficiency does not merely slow down convergence but can fundamentally bias the optimization by trapping the algorithm in metastable states. 

We now demonstrate the advantages of our sampling strategy in full variational Monte Carlo optimizations. In frustrated systems, inefficient sampling can introduce systematic bias, trapping the optimization in metastable local minima.
To probe this, we fix a single disorder realization of the QSG and perform VMC optimization using the three sampling strategies.
We employ identical hyperparameters ($D=3, \chi=9$, learning rate $0.1$, batch size 1000 ) and stochastic reconfiguration (SR)~\cite{SR_1,SR_2,SR_3} for 5000 steps. Statistical reliability is assessed via independent runs (100 seeds for $L=10$ and 20 seeds for $L=16$).
% . All three sampling strategies are initialized from the same PEPS tensor configuration and optimized using identical hyperparameters: bond dimension $D=3$, environment bond dimension $\chi=9$, learning rate $0.1$, a batch size of 1000 Markov chains, and stochastic reconfiguration (SR)~\cite{SR_1,SR_2,SR_3} for gradient evaluation. Each optimization is run for 5000 steps. To assess statistical robustness, we repeat this procedure using different random number seeds. For system size $L=10$, we perform 100 independent runs, while for $L=16$, we perform 20 runs. At each optimization step, we compute the mean energy and variance across all seeds.
The results are shown in Fig.~\ref{fig:spin_glass_energy}. For both system sizes, the Metropolis-based optimization converges to significantly higher energies with large variance across seeds, indicating frequent trapping in local minima. In contrast, the autoregressive and the hybrid strategies achieve substantially lower average energies with minimal variance. This behavior demonstrates that non-local sampling not only accelerates convergence but also stabilizes the optimization trajectory by enabling consistent exploration of the variational landscape.

\begin{figure}
    \centering
    \includegraphics[width=1.0\linewidth]{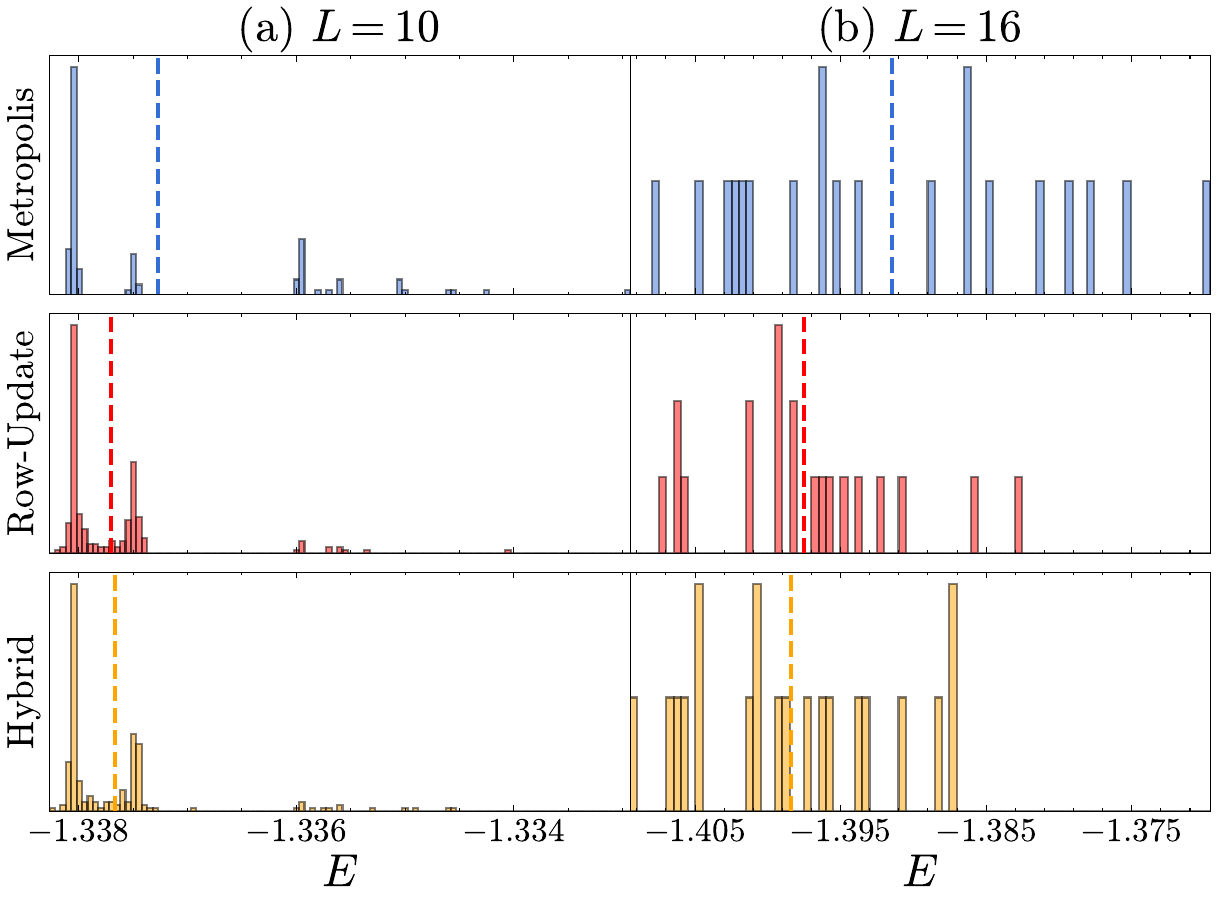}
    \caption{Distribution of variational energies obtained after 5000 optimization steps for the QSG model: (a) $L=10$ and (b) $L=16$ systems.
    Histograms show the final energies from independent optimization runs, with the vertical dashed line indicating the mean value.
    Compared to the Metropolis sampler, the autoregressive row-update and hybrid methods produce narrower energy distributions with lower mean and minimum energies.}
    \label{fig:energy_distribution}
\end{figure}

To further characterize the statistical behavior of the optimization outcomes, 
Fig.~\ref{fig:energy_distribution} shows the distributions of variational energies obtained after 5000 optimization steps across independent runs.
For both system sizes, the Metropolis sampler produces a broad distribution extending toward relatively high energies, reflecting a strong sensitivity to stochastic sampling trajectories and frequent trapping in metastable configurations.
In comparison, the autoregressive row-update method leads to a visibly narrower distribution and systematically lower mean energies.
The hybrid strategy further reduces the spread of the distribution and achieves the lowest energies among the three methods.
These results suggest that incorporating non-local row-wise updates can alleviate, though not completely eliminate, optimization instability induced by purely local Metropolis sampling in frustrated quantum systems.

\emph{Conclusion.}---In this work, we have explored a non-local autoregressive row-wise sampling strategy for projected entangled-pair state (PEPS) based variational Monte Carlo simulations. 
By exploiting efficient single-layer tensor network contractions, the proposed method evaluates exact conditional probabilities along a given row and enables direct sampling of entire spin rows conditioned on their environment. 
This realizes collective Monte Carlo updates that extend beyond strictly local Metropolis moves while remaining comparable in computational cost.

These results indicate that autoregressive row sampling provides a useful enhancement to existing PEPS-VMC methodologies, particularly in regimes where local updates alone become inefficient. 
In contrast to neural-network quantum states, where autoregressive sampling is intrinsic to the ansatz~\cite{wu_solving_2019}, our work shows that structured tensor-network states can also benefit significantly from carefully designed non-local sampling schemes, suggesting that the potential of PEPS-based VMC has not yet been fully explored.

Looking ahead, several extensions of the present framework are worth exploring. 
In particular, incorporating explicit constraints to enforce conserved quantities would enable applications to particle-number–conserving spin and fermionic models, paving the way for studies of the doped 2D Hubbard model.
Beyond ground-state optimization, the row-update paradigm may also be adapted to real-time schemes~\cite{wu2025}, as well as to isometric or gauge-fixed PEPS formulations~\cite{IsoPEPS,IsoPEPS-Wu}, where improved sampling of intermediate configurations could enhance numerical stability. 
More broadly, these possibilities suggest that autoregressive sampling can serve as a flexible component in future tensor-network approaches to strongly correlated and frustrated quantum systems.

\begin{acknowledgments}
    This work is supported by Projects 12325501, 12247104, 12275263, and 12405047 of the National Natural Science Foundation of China, and the Innovation Program for Quantum Science and Technology (under Grant No. 2021ZD0301900). YD thanks the support from the Natural Science Foundation of Fujian Province of China (under Grant No. 2023J02032).
    Y.W. is supported by a start-up grant from IOP-CAS.
\end{acknowledgments}

\vskip 1cm    

% \section{B}
\bibliography{references.bib}

@misc{wu2025algorithms,
      title={Algorithms for variational Monte Carlo calculations of fermion projected entangled pair states in the swap gates formulation and the detailed balance of tensor network sequential sampling}, 
      author={Yantao Wu and Zhehao Dai},
      year={2025},
      eprint={2506.20106},
      archivePrefix={arXiv},
      primaryClass={cond-mat.str-el},
      url={https://arxiv.org/abs/2506.20106}, 
}

@article{PhysRevLett.69.2863,
  title = {Density matrix formulation for quantum renormalization groups},
  author = {White, Steven R.},
  journal = {Phys. Rev. Lett.},
  volume = {69},
  issue = {19},
  pages = {2863--2866},
  numpages = {0},
  year = {1992},
  month = {Nov},
  publisher = {American Physical Society},
  doi = {10.1103/PhysRevLett.69.2863},
  url = {https://link.aps.org/doi/10.1103/PhysRevLett.69.2863}
}

@article{PhysRevLett.93.040502,
  title = {Efficient Simulation of One-Dimensional Quantum Many-Body Systems},
  author = {Vidal, Guifr\'e},
  journal = {Phys. Rev. Lett.},
  volume = {93},
  issue = {4},
  pages = {040502},
  numpages = {4},
  year = {2004},
  month = {Jul},
  publisher = {American Physical Society},
  doi = {10.1103/PhysRevLett.93.040502},
  url = {https://link.aps.org/doi/10.1103/PhysRevLett.93.040502}
}

@article{PhysRevLett.93.227205,
  title = {Density Matrix Renormalization Group and Periodic Boundary Conditions: A Quantum Information Perspective},
  author = {Verstraete, F. and Porras, D. and Cirac, J. I.},
  journal = {Phys. Rev. Lett.},
  volume = {93},
  issue = {22},
  pages = {227205},
  numpages = {4},
  year = {2004},
  month = {Nov},
  publisher = {American Physical Society},
  doi = {10.1103/PhysRevLett.93.227205},
  url = {https://link.aps.org/doi/10.1103/PhysRevLett.93.227205}
}

@article{PhysRevLett.96.220601,
  title = {Criticality, the Area Law, and the Computational Power of Projected Entangled Pair States},
  author = {Verstraete, F. and Wolf, M. M. and Perez-Garcia, D. and Cirac, J. I.},
  journal = {Phys. Rev. Lett.},
  volume = {96},
  issue = {22},
  pages = {220601},
  numpages = {4},
  year = {2006},
  month = {Jun},
  publisher = {American Physical Society},
  doi = {10.1103/PhysRevLett.96.220601},
  url = {https://link.aps.org/doi/10.1103/PhysRevLett.96.220601}
}

@article{PhysRevLett.99.220405,
  author = {Vidal, G.},
  journal = {Phys. Rev. Lett.},
  volume = {99},
  issue = {22},
  pages = {220405},
  numpages = {4},
  year = {2007},
  month = {Nov},
  publisher = {American Physical Society},
  doi = {10.1103/PhysRevLett.99.220405},
  url = {https://link.aps.org/doi/10.1103/PhysRevLett.99.220405}
}

@article{PhysRevLett.101.250602,
  title = {Classical {Simulation} of {Infinite-Size} {Quantum} {Lattice} {Systems} in {Two} {Spatial} {Dimensions}},
  author = {Jordan, J. and Or\'us, R. and Vidal, G. and Verstraete, F. and Cirac, J. I.},
  journal = {Phys. Rev. Lett.},
  volume = {101},
  issue = {25},
  pages = {250602},
  numpages = {4},
  year = {2008},
  month = {Dec},
  publisher = {American Physical Society},
  doi = {10.1103/PhysRevLett.101.250602},
  url = {https://link.aps.org/doi/10.1103/PhysRevLett.101.250602}
}

@article{PhysRevLett.101.090603,
  title = {Accurate Determination of Tensor Network State of Quantum Lattice Models in Two Dimensions},
  author = {Jiang, H. C. and Weng, Z. Y. and Xiang, T.},
  journal = {Phys. Rev. Lett.},
  volume = {101},
  issue = {9},
  pages = {090603},
  numpages = {4},
  year = {2008},
  month = {Aug},
  publisher = {American Physical Society},
  doi = {10.1103/PhysRevLett.101.090603},
  url = {https://link.aps.org/doi/10.1103/PhysRevLett.101.090603}
}

@article{PhysRevB.90.064425,
  title = {Algorithms for finite projected entangled pair states},
  author = {Lubasch, Michael and Cirac, J. Ignacio and Ba\~nuls, Mari-Carmen},
  journal = {Phys. Rev. B},
  volume = {90},
  issue = {6},
  pages = {064425},
  numpages = {16},
  year = {2014},
  month = {Aug},
  publisher = {American Physical Society},
  doi = {10.1103/PhysRevB.90.064425},
  url = {https://link.aps.org/doi/10.1103/PhysRevB.90.064425}
}

@article{PhysRevB.95.195154,
  title = {Gradient optimization of finite projected entangled pair states},
  author = {Liu, Wen-Yuan and Dong, Shao-Jun and Han, Yong-Jian and Guo, Guang-Can and He, Lixin},
  journal = {Phys. Rev. B},
  volume = {95},
  issue = {19},
  pages = {195154},
  numpages = {8},
  year = {2017},
  month = {May},
  publisher = {American Physical Society},
  doi = {10.1103/PhysRevB.95.195154},
  url = {https://link.aps.org/doi/10.1103/PhysRevB.95.195154}
}

@article{PhysRevB.99.195153,
  title = {Gradient optimization of fermionic projected entangled pair states on directed lattices},
  author = {Dong, Shao-Jun and Wang, Chao and Han, Yongjian and Guo, Guang-can and He, Lixin},
  journal = {Phys. Rev. B},
  volume = {99},
  issue = {19},
  pages = {195153},
  numpages = {10},
  year = {2019},
  month = {May},
  publisher = {American Physical Society},
  doi = {10.1103/PhysRevB.99.195153},
  url = {https://link.aps.org/doi/10.1103/PhysRevB.99.195153}
}

@article{PhysRevLett.124.037201,
  title = {Isometric Tensor Network States in Two Dimensions},
  author = {Zaletel, Michael P. and Pollmann, Frank},
  journal = {Phys. Rev. Lett.},
  volume = {124},
  issue = {3},
  pages = {037201},
  numpages = {5},
  year = {2020},
  month = {Jan},
  publisher = {American Physical Society},
  doi = {10.1103/PhysRevLett.124.037201},
  url = {https://link.aps.org/doi/10.1103/PhysRevLett.124.037201}
}

@article{PhysRevB.103.235155,
  title = {Accurate simulation for finite projected entangled pair states in two dimensions},
  author = {Liu, Wen-Yuan and Huang, Yi-Zhen and Gong, Shou-Shu and Gu, Zheng-Cheng},
  journal = {Phys. Rev. B},
  volume = {103},
  issue = {23},
  pages = {235155},
  numpages = {13},
  year = {2021},
  month = {Jun},
  publisher = {American Physical Society},
  doi = {10.1103/PhysRevB.103.235155},
  url = {https://link.aps.org/doi/10.1103/PhysRevB.103.235155}
}

@article{SCHOLLWOCK201196,
title = {The density-matrix renormalization group in the age of matrix product states},
journal = {Annals of Physics},
volume = {326},
number = {1},
pages = {96-192},
year = {2011},
note = {January 2011 Special Issue},
issn = {0003-4916},
doi = {https://doi.org/10.1016/j.aop.2010.09.012},
url = {https://www.sciencedirect.com/science/article/pii/S0003491610001752},
author = {Ulrich Schollwöck},
}

@article{PhysRevB.48.10345,
  title = {Density-matrix algorithms for quantum renormalization groups},
  author = {White, Steven R.},
  journal = {Phys. Rev. B},
  volume = {48},
  issue = {14},
  pages = {10345--10356},
  numpages = {0},
  year = {1993},
  month = {Oct},
  publisher = {American Physical Society},
  doi = {10.1103/PhysRevB.48.10345},
  url = {https://link.aps.org/doi/10.1103/PhysRevB.48.10345}
}

@article{PhysRevB.80.094403,
  title = {Simulation of two-dimensional quantum systems on an infinite lattice revisited: Corner transfer matrix for tensor contraction},
  author = {Or\'us, Rom\'an and Vidal, Guifr\'e},
  journal = {Phys. Rev. B},
  volume = {80},
  issue = {9},
  pages = {094403},
  numpages = {4},
  year = {2009},
  month = {Sep},
  publisher = {American Physical Society},
  doi = {10.1103/PhysRevB.80.094403},
  url = {https://link.aps.org/doi/10.1103/PhysRevB.80.094403}
}

@article{PhysRevLett.103.160601,
  title = {Second Renormalization of Tensor-Network States},
  author = {Xie, Z. Y. and Jiang, H. C. and Chen, Q. N. and Weng, Z. Y. and Xiang, T.},
  journal = {Phys. Rev. Lett.},
  volume = {103},
  issue = {16},
  pages = {160601},
  numpages = {4},
  year = {2009},
  month = {Oct},
  publisher = {American Physical Society},
  doi = {10.1103/PhysRevLett.103.160601},
  url = {https://link.aps.org/doi/10.1103/PhysRevLett.103.160601}
}

@article{PhysRevB.86.045139,
  title = {Coarse-graining renormalization by higher-order singular value decomposition},
  author = {Xie, Z. Y. and Chen, J. and Qin, M. P. and Zhu, J. W. and Yang, L. P. and Xiang, T.},
  journal = {Phys. Rev. B},
  volume = {86},
  issue = {4},
  pages = {045139},
  numpages = {9},
  year = {2012},
  month = {Jul},
  publisher = {American Physical Society},
  doi = {10.1103/PhysRevB.86.045139},
  url = {https://link.aps.org/doi/10.1103/PhysRevB.86.045139}
}

@article{PhysRevB.92.201111,
  title = {Excitations and the tangent space of projected entangled-pair states},
  author = {Vanderstraeten, Laurens and Mari\"en, Micha\"el and Verstraete, Frank and Haegeman, Jutho},
  journal = {Phys. Rev. B},
  volume = {92},
  issue = {20},
  pages = {201111},
  numpages = {5},
  year = {2015},
  month = {Nov},
  publisher = {American Physical Society},
  doi = {10.1103/PhysRevB.92.201111},
  url = {https://link.aps.org/doi/10.1103/PhysRevB.92.201111}
}

@article{PhysRevLett.134.256502,
  title = {Accurate Simulation of the Hubbard Model with Finite Fermionic Projected Entangled Pair States},
  author = {Liu, Wen-Yuan and Zhai, Huanchen and Peng, Ruojing and Gu, Zheng-Cheng and Chan, Garnet Kin-Lic},
  journal = {Phys. Rev. Lett.},
  volume = {134},
  issue = {25},
  pages = {256502},
  numpages = {8},
  year = {2025},
  month = {Jun},
  publisher = {American Physical Society},
  doi = {10.1103/r4q9-4yvj},
  url = {https://link.aps.org/doi/10.1103/r4q9-4yvj}
}

@article{PhysRevLett.99.220602,
  title = {Variational Quantum Monte Carlo Simulations with Tensor-Network States},
  author = {Sandvik, A. W. and Vidal, G.},
  journal = {Phys. Rev. Lett.},
  volume = {99},
  issue = {22},
  pages = {220602},
  numpages = {4},
  year = {2007},
  month = {Nov},
  publisher = {American Physical Society},
  doi = {10.1103/PhysRevLett.99.220602},
  url = {https://link.aps.org/doi/10.1103/PhysRevLett.99.220602}
}

@article{PhysRevLett.100.040501,
  title = {Simulation of Quantum Many-Body Systems with Strings of Operators and Monte Carlo Tensor Contractions},
  author = {Schuch, Norbert and Wolf, Michael M. and Verstraete, Frank and Cirac, J. Ignacio},
  journal = {Phys. Rev. Lett.},
  volume = {100},
  issue = {4},
  pages = {040501},
  numpages = {4},
  year = {2008},
  month = {Jan},
  publisher = {American Physical Society},
  doi = {10.1103/PhysRevLett.100.040501},
  url = {https://link.aps.org/doi/10.1103/PhysRevLett.100.040501}
}

@article{Metropolis1953EquationOS,
  title   = {Equation of state calculations by fast computing machines},
  author  = {N. Metropolis and Arianna W. Rosenbluth and Marshall N. Rosenbluth and A. H. Teller and Edward Teller},
  journal = {J. Chem. Phys.},
  year    = {1953},
  volume  = {21},
  pages   = {1087-1092}
}

@article{PhysRev.145.224,
  title = {Diffusion Constants near the Critical Point for Time-Dependent Ising Models. I},
  author = {Kawasaki, Kyozi},
  journal = {Phys. Rev.},
  volume = {145},
  issue = {1},
  pages = {224--230},
  numpages = {0},
  year = {1966},
  month = {May},
  publisher = {American Physical Society},
  doi = {10.1103/PhysRev.145.224},
  url = {https://link.aps.org/doi/10.1103/PhysRev.145.224}
}

@article{Edwards1975,
author = {Edwards, S F and Anderson, P W},
doi = {10.1088/0305-4608/5/5/017},
issn = {0305-4608},
journal = {J. Phys. F: Met. Phys.},
month = {may},
number = {5},
pages = {965--974},
title = {{Theory of spin glasses}},
url = {https://iopscience.iop.org/article/10.1088/0305-4608/5/5/017/pdf https://iopscience.iop.org/article/10.1088/0305-4608/5/5/017},
volume = {5},
year = {1975}
}

@article{binder1986,
  title = {Spin glasses: Experimental facts, theoretical concepts, and open questions},
  author = {Binder, K. and Young, A. P.},
  journal = {Rev. Mod. Phys.},
  volume = {58},
  issue = {4},
  pages = {801--976},
  numpages = {0},
  year = {1986},
  month = {Oct},
  publisher = {American Physical Society},
  doi = {10.1103/RevModPhys.58.801},
  url = {https://link.aps.org/doi/10.1103/RevModPhys.58.801}
}

@Article{SciPostPhys.14.5.123,
	title={{Collective Monte Carlo updates through tensor network renormalization}},
	author={Miguel Frías-Pérez and Michael Mariën and David Pérez García and Mari Carmen Bañuls and Sofyan Iblisdir},
	journal={SciPost Phys.},
	volume={14},
	pages={123},
	year={2023},
	publisher={SciPost},
	doi={10.21468/SciPostPhys.14.5.123},
	url={https://scipost.org/10.21468/SciPostPhys.14.5.123},
}

@article{PhysRevB.111.094201,
  title = {Tensor network Monte Carlo simulations for the two-dimensional random-bond Ising model},
  author = {Chen, Tao and Guo, Erdong and Zhang, Wanzhou and Zhang, Pan and Deng, Youjin},
  journal = {Phys. Rev. B},
  volume = {111},
  issue = {9},
  pages = {094201},
  numpages = {14},
  year = {2025},
  month = {Mar},
  publisher = {American Physical Society},
  doi = {10.1103/PhysRevB.111.094201},
  url = {https://link.aps.org/doi/10.1103/PhysRevB.111.094201}
}

@misc{chen2025batchtnmc,
      title={BatchTNMC: Efficient sampling of two-dimensional spin glasses using tensor network Monte Carlo}, 
      author={Tao Chen and Jingtong Zhang and Jing Liu and Youjin Deng and Pan Zhang},
      year={2025},
      eprint={2509.19006},
      archivePrefix={arXiv},
      primaryClass={cond-mat.stat-mech},
      url={https://arxiv.org/abs/2509.19006}, 
}

@misc{chen2025tensornetworkmarkovchain,
      title={Tensor Network Markov Chain Monte Carlo: Efficient Sampling of Three-Dimensional Spin Glasses and Beyond}, 
      author={Tao Chen and Jing Liu and Youjin Deng and Pan Zhang},
      year={2025},
      eprint={2509.23945},
      archivePrefix={arXiv},
      primaryClass={cond-mat.stat-mech},
      url={https://arxiv.org/abs/2509.23945}, 
}

@article{TFIM_deng,
  title = {Cluster Monte Carlo simulation of the transverse Ising model},
  author = {Bl\"ote, Henk W. J. and Deng, Youjin},
  journal = {Phys. Rev. E},
  volume = {66},
  issue = {6},
  pages = {066110},
  numpages = {8},
  year = {2002},
  month = {Dec},
  publisher = {American Physical Society},
  doi = {10.1103/PhysRevE.66.066110},
  url = {https://link.aps.org/doi/10.1103/PhysRevE.66.066110}
}

@article{Bernaschi2024,
  author = {Massimo Bernaschi and Isidoro González-Adalid Pemartín and Víctor Martín-Mayor and Giorgio Parisi},
  title = {The quantum transition of the two-dimensional Ising spin glass},
  journal = {Nature},
  volume = {631},
  number = {8022},
  pages = {749--754},
  year = {2024},
  doi = {10.1038/s41586-024-07647-y},
  url = {https://doi.org/10.1038/s41586-024-07647-y}
}

@article{SR_1,
  title = {Green Function Monte Carlo with Stochastic Reconfiguration},
  author = {Sorella, Sandro},
  journal = {Phys. Rev. Lett.},
  volume = {80},
  issue = {20},
  pages = {4558--4561},
  numpages = {0},
  year = {1998},
  month = {May},
  publisher = {American Physical Society},
  doi = {10.1103/PhysRevLett.80.4558},
  url = {https://link.aps.org/doi/10.1103/PhysRevLett.80.4558}
}

@article{SR_2,
  title = {Optimizing large parameter sets in variational quantum Monte Carlo},
  author = {Neuscamman, Eric and Umrigar, C. J. and Chan, Garnet Kin-Lic},
  journal = {Phys. Rev. B},
  volume = {85},
  issue = {4},
  pages = {045103},
  numpages = {6},
  year = {2012},
  month = {Jan},
  publisher = {American Physical Society},
  doi = {10.1103/PhysRevB.85.045103},
  url = {https://link.aps.org/doi/10.1103/PhysRevB.85.045103}
}

@article{SR_3,
  title = {Direct sampling of projected entangled-pair states},
  author = {Vieijra, Tom and Haegeman, Jutho and Verstraete, Frank and Vanderstraeten, Laurens},
  journal = {Phys. Rev. B},
  volume = {104},
  issue = {23},
  pages = {235141},
  numpages = {17},
  year = {2021},
  month = {Dec},
  publisher = {American Physical Society},
  doi = {10.1103/PhysRevB.104.235141},
  url = {https://link.aps.org/doi/10.1103/PhysRevB.104.235141}
}

@inbook{Becca_Sorella_2017, 
    place={Cambridge}, 
    title={Variational Monte Carlo}, 
    booktitle={Quantum Monte Carlo Approaches for Correlated Systems}, 
    publisher={Cambridge University Press}, 
    author={Becca, Federico and Sorella, Sandro}, 
    year={2017}, 
    pages={101–102}
}

@article{wu_solving_2019,
	title = {Solving {Statistical} {Mechanics} {Using} {Variational} {Autoregressive} {Networks}},
	volume = {122},
	issn = {0031-9007, 1079-7114},
	url = {https://link.aps.org/doi/10.1103/PhysRevLett.122.080602},
	doi = {10.1103/PhysRevLett.122.080602},
	number = {8},
	urldate = {2024-07-04},
	journal = {Physical Review Letters},
	author = {Wu, Dian and Wang, Lei and Zhang, Pan},
	month = {feb},
	year = {2019},
	pages = {080602},
}

@misc{wu2025,
      title={Real-Time Dynamics in Two Dimensions with Tensor Network States via Time-Dependent Variational Monte Carlo}, 
      author={Yantao Wu},
      year={2025},
      eprint={2512.06768},
      archivePrefix={arXiv},
      primaryClass={cond-mat.str-el},
      url={https://arxiv.org/abs/2512.06768}, 
}

@article{IsoPEPS,
  title = {Dual-Isometric Projected Entangled Pair States},
  author = {Yu, Xie-Hang and Cirac, J. Ignacio and Kos, Pavel and Styliaris, Georgios},
  journal = {Phys. Rev. Lett.},
  volume = {133},
  issue = {19},
  pages = {190401},
  numpages = {7},
  year = {2024},
  month = {Nov},
  publisher = {American Physical Society},
  doi = {10.1103/PhysRevLett.133.190401},
  url = {https://link.aps.org/doi/10.1103/PhysRevLett.133.190401}
}

@article{IsoPEPS-Wu,
  title = {Alternating and Gaussian Fermionic Isometric Tensor Network States},
  author = {Wu, Yantao and Dai, Zhehao and Anand, Sajant and Lin, Sheng-Hsuan and Yang, Qi and Wang, Lei and Pollmann, Frank and Zaletel, Michael P.},
  journal = {PRX Quantum},
  volume = {6},
  issue = {4},
  pages = {040324},
  numpages = {20},
  year = {2025},
  month = {Nov},
  publisher = {American Physical Society},
  doi = {10.1103/8ypw-c8t4},
  url = {https://link.aps.org/doi/10.1103/8ypw-c8t4}
}
\end{document}

\title{Supplemental Material for ``Variational Monte Carlo (VMC) with row-update Projected Entangled-Pair States (PEPS) and its applications in quantum spin glasses''}
\maketitle
% \section*{Appendix: Algorithm and Implementation Details}

\section{Single-Layer Autoregressive Row-Update Strategy}
The central ingredient of our approach is the efficient evaluation of conditional probabilities for an entire row of spins given its surrounding environment. 
Building upon the sequential update framework introduced by Liu \textit{et al.}~\cite{PhysRevB.95.195154,PhysRevB.103.235155}, we implement a non-local update based on an autoregressive sampling scheme for the row configuration $\s_y = (s_{y,1}, \dots, s_{y,L})$.
Let $\s_{\setminus y}$ denote the configuration of all spins outside row $y$, whose influence is encoded in the top boundary MPS $\Phi_{y-1}$ and the approximate bottom environment $\Env_{y+1}$. 
The conditional probability of the row configuration can be factorized according to the chain rule,
\begin{equation}
    P(\s_y | \s_{\setminus y}) = \prod_{i=1}^L P(s_{y,i} | \s_{y,<i}, \s_{\setminus y}),
    \label{eq:autoregressive_prob}
\end{equation}
which enables sequential sampling of spins along the row.

\begin{figure}
    \centering
    \includegraphics[width=1.0\linewidth]{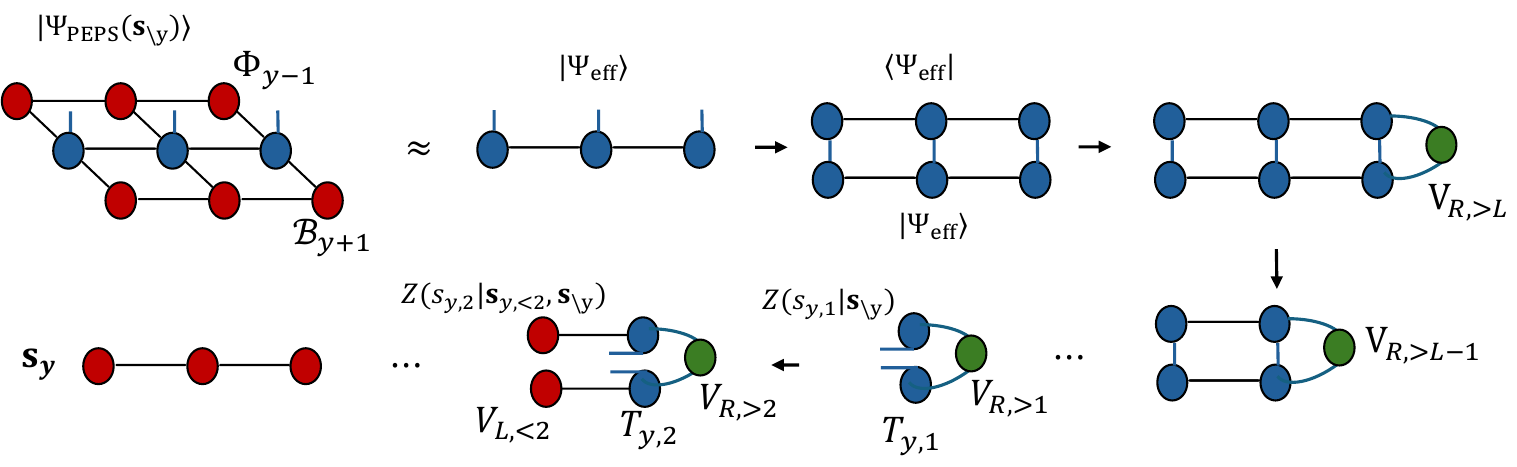}
    \caption{Schematic illustration of the row-wise autoregressive sampling procedure.}
    \label{fig:RowUpdate}
\end{figure}

In practice, the conditional distribution defined in Eq.~(\ref{eq:autoregressive_prob}) is represented by an effective one-dimensional state $\PsiEff$, formulated as a matrix product state (MPS). 
This effective MPS is obtained by contracting the top boundary $\Phi_{y-1}$, the PEPS tensors on row $y$, and the bottom environment $\Env_{y+1}$, followed by a truncation to a maximum bond dimension $\chi$, as illustrated in Fig.~\ref{fig:RowUpdate}. 
Once $\PsiEff$ is constructed, spins along the row are sampled sequentially in a rejection-free manner using standard autoregressive sampling of MPS.

This procedure involves two sweeps:
\begin{enumerate}
    \item \textit{Environment construction (right to left):}
    A preparatory sweep from right to left ($i=L$ to $2$) is performed to construct the right environments $V_{R,>i}$.
    In this sweep, the local tensors of the effective MPS $\Psi_{\mathrm{eff}}$ and their conjugates are successively contracted, and the resulting partial contractions are stored as right environment tensors.

    \item \textit{Sequential sampling (left to right):}
    The sampling then proceeds sequentially from left to right ($i=1$ to $L$).
    A left environment $V_{L,<i}$ is progressively built from the already sampled spins.
    At each site $i$, the local tensor $T_{y,i}$ (and its conjugate) is contracted with the left environment $V_{L,<i}$ and the precomputed right environment $V_{R,>i}$, yielding a matrix proportional to the local marginal weight
    $Z(s_{y,i} \mid \s_{y,<i}, \s_{\setminus y})$.
    The diagonal elements of this matrix define the conditional probability
    $P(s_{y,i} \mid \s_{y,<i}, \s_{\setminus y})$, from which the spin $s_{y,i}$ is directly sampled.
    The left environment is then updated by absorbing the sampled tensor $T_{y,i}^{s_{y,i}}$, and the procedure is iterated along the row.
\end{enumerate}

This row-wise autoregressive update constitutes a collective, non-local Monte Carlo move within the PEPS-VMC framework. 
Conceptually, it is closely related to recent developments in tensor-network Monte Carlo methods for classical systems~\cite{SciPostPhys.14.5.123,PhysRevB.111.094201,chen2025tensornetworkmarkovchain,chen2025batchtnmc}.

\section{Pseudocode}

The overall VMC optimization procedure is outlined in Algorithm~\ref{alg:vmc_main}. 
In each optimization step, we first update the set of bottom environment approximations $\{\mathcal{B}_y\}$ based on the current tensors $T$. 
We then invoke the row-update subroutine to generate a batch of samples $\s_{\text{batch}}$. 
The specific autoregressive row-sampling procedure is detailed in Algorithm~\ref{alg:row_update}. 
Gradients are subsequently computed by averaging over this batch, followed by a stochastic reconfiguration (SR) update of the tensor parameters.
% ---------------------------------------------------------
% Algorithm 1: Main VMC Loop (SR Only)
% ---------------------------------------------------------
\begin{algorithm}[H]
\caption{VMC-PEPS Optimization}
\label{alg:vmc_main}
\begin{algorithmic}[1]
\Procedure{VMC-PEPS}{$T$} \Comment{Input PEPS tensors $T$}
    \State Initialize parameters $T$ randomly
    \While{$i \le I_{\max}$} \Comment{Optimization Loop}
        
        \State \textbf{1. Environment Pre-calculation:}
        \State Update Bottom Environments $\{\Env_y\}_{y=2}^{L+1}$ using Boundary MPS method
        
        \State \textbf{2. Batch Sampling:}
        \State $\s_{\text{batch}} \leftarrow$ \Call{RowUpdateSampling}{$T, \{\Env_y\}, M$} 
        \Comment{Get $M$ samples}

        \State \textbf{3. Statistics Accumulation (Batch):}
        \State $E_{\text{tot}} \leftarrow 0, \quad P_1 \leftarrow 0, \quad P_2 \leftarrow 0$
        \State $\mathcal{S}_{\text{matrix}} \leftarrow 0$ \Comment{Accumulator for SR Covariance Matrix}
        
        \ForAll{$\s_m \in \s_{\text{batch}}$} \Comment{Parallel reduction over batch}
            \State $\Psi(\s_m) \leftarrow$ ContractNetwork($\s_m$)
            
            \State \Comment{Compute Local Energy $E_{\text{loc}}$}
            \State  Compute non-zero diagonal elements $\langle \s'_n|H|\s_m\rangle \neq 0$
            \State $E_{\text{loc}}(\s_m) \leftarrow \sum_n \frac{\Psi(\s'_n)}{\Psi(\s_m)} \langle \s'_n|H|\s_m\rangle$
            
            \State \Comment{Compute Log-Derivatives $A^{s_k}$}
            \State $A^{s_k}(\s_m) \leftarrow$ ContractDefectNetwork($\s_m, k$)
            
            \State \Comment{Accumulate Gradients and SR Matrix elements}
            \State $E_{\text{tot}} \leftarrow E_{\text{tot}} + E_{\text{loc}}(\s_m)$
            \State $P_1 \leftarrow P_1 + A^{s_k}(\s_m) \cdot E_{\text{loc}}(\s_m)$
            \State $P_2 \leftarrow P_2 + A^{s_k}(\s_m)$
            \State $\mathcal{S}_{\text{matrix}} \leftarrow \mathcal{S}_{\text{matrix}} + A^{s_k}(\s_m) \cdot (A^{s_k}(\s_m))^T$ \Comment{Outer product}
        \EndFor

        \State \textbf{4. Stochastic Reconfiguration Update:}
        \State $E_{\text{mean}} \leftarrow E_{\text{tot}}/M$
        \State $\bar{g} \leftarrow P_1/M - (P_2/M) \cdot E_{\text{mean}}$ \Comment{Energy Gradient}
        \State $\mathcal{S} \leftarrow \mathcal{S}_{\text{matrix}}/M - (P_2/M) \cdot (P_2/M)^T$ \Comment{Covariance Matrix}
        
        \State \Comment{Solve linear system $\mathcal{S} x = \bar{g}$ for SR}
        \State $\delta T \leftarrow \text{LinearSolve}(\mathcal{S} + \epsilon I, \bar{g})$ \Comment{Regularization $\epsilon$ if needed}
        \State $T \leftarrow T - \eta \cdot \delta T$ \Comment{Update parameters}
    \EndWhile
\EndProcedure
\end{algorithmic}
\end{algorithm}

% ---------------------------------------------------------
% Algorithm 2: Row-Update Subroutine
% ---------------------------------------------------------
\begin{algorithm}[H]
\caption{Batch Autoregressive Row-Update Sampling}
\label{alg:row_update}
\begin{algorithmic}[1]
\Procedure{RowUpdateSampling}{$T, \{\Env_y\}, M$}
    \State $\s_{\text{batch}} \leftarrow$ Initialize batch of size $M$ with empty rows
    \State $\Phi_{0} \leftarrow$ Initialize Vacuum Top Boundary (Batch)
    
    \For{$y \leftarrow 1$ to $L$} \Comment{Sweep from Top to Bottom}
        \State Load Bottom Environment $\Env_{y+1}$ \Comment{Pre-calculated}
        
        \State \textbf{Construct Effective MPS:}
        \State $\PsiEff \leftarrow \text{Contract}(\Phi_{y-1}, T_{y}, \Env_{y+1})$
        \State $\PsiEff \leftarrow \text{Truncate}(\PsiEff, \chi)$ \Comment{SVD Truncation}
        
        \State \textbf{Autoregressive Sampling (Batch):}
        \State Contract $\PsiEff$ from right to left to construct the right environments $V_{R,>i}$
        \State $\s_y \leftarrow$ Empty row configurations for batch
        \For{$x \leftarrow 1$ to $L$} \Comment{Left to Right}
            \State Compute $P(s_{y,x} | \s_{y,<x}, \s_{\setminus y})$
            \State Sample $s_{y,x} \sim P(s_{y,x})$ for all $m \in [1, M]$
            \State Append $s_{y,x}$ to $\s_y$
        \EndFor
        \State Append $\s_y$ to corresponding $\s_{\text{batch}}$
        
        \State \textbf{Update Top Boundary:}
        \State $\Phi_{y} \leftarrow \text{Contract}(\Phi_{y-1}, T_y[\s_y])$ \Comment{Propagate Top Env}
    \EndFor
    
    \State \Return $\s_{\text{batch}}$
\EndProcedure
\end{algorithmic}
\end{algorithm}

\vskip 1cm    
\bibliography{references.bib}